\documentclass[%
 reprint,
 amsmath,amssymb,
 aps,
]{revtex4-2}

\usepackage{graphicx}% Include figure files
\usepackage{dcolumn}% Align table columns on decimal point
\usepackage{bm}% bold math
\usepackage{xcolor}
\usepackage{bm}
\usepackage[english]{babel}
\usepackage[autostyle]{csquotes}
\usepackage{lineno}
\begin{document}
%\linenumbers
%\preprint{APS/123-QED}

\title{
Colliding waves lead to enhanced  absorption of high intensity laser pulses in plasma micro-globules 
%OR
%Electron waves and currents collide to enhance intense laser absorption in plasma microglobules 
%OR 
%Boosted laser absorption by colliding electron waves and currents in a microglobule plasma
%OR
%Laser energy absorption gets boosted by colliding electron waves and currents in a micro-globule plasma target
%OR 
%Efficient heating of micro globular plasma targets by colliding waves at sub relativistic laser intensity
%OR 
%Colliding waves aid microglobular plasma targets to efficiently generate energetic electrons by sub-relativistic intensity lasers. 
%OR Efficient electron heating by laser in finite sized plasma micro-globular targets by  repeated collisions of surface and bulk waves 
%OR
%Efficient energetic electron generation by sub-relativistic laser intensity by colliding waves in micro-globular plasma 
}
%\title{{Surface Oscillations in Magnetized Overdense Plasma: Unleashing the Formation of Lower Hybrid Waves and Transverse Filaments}}
%\title{Laser-Induced Surface Oscillations: Unleashing the Formation of Lower Hybrid waves, and Filaments in Magnetized Overdense Plasma\\
%\textcolor{blue}{Surface Oscillations in Magnetized Overdense Plasma: Unleashing the Formation of Lower Hybrid Waves and Transverse Filaments}}

% Force line breaks with \\
\author{Animesh Sharma}
\email{animesh.sharma.research@gmail.com} 
\author{Amita Das}%
 \email{amitadas3@yahoo.com}
%\author{...}
\affiliation{%
 Department of Physics, 
 Indian Institute of Technology Delhi,\\
 Hauz Khas, New Delhi-110016, India
}%
\author{G. Ravindra Kumar}
\affiliation{%
Tata Institute of Fundamental Research\\
Colaba, Mumbai\\
}%
%\affiliation{
%\affiliation{%
\date{\today}% It is always \today, today,
\begin{abstract}
%A new mechanism of enhanced laser energy absorption in plasma microglobules is demonstrated with the help of two-dimensional Particle-In-Cell (PIC) simulations.  The mechanism relies on the excitation of surface and bulk waves and the occurrence of repeated collisions in the confines of the finite-sized microglobular target. The episodic increase in the average particle energy correlates with the repeated collision of the surface and bulk waves that get excited by the laser on the target.  It is shown that the size of the microglobular target governs the efficiency of absorption and the timings of episodic events of energy enhancement.    This study thereby illustrates the novel efficient possibility that a closed plasma target provides for energy extraction.  Parallels of such colliding waves creating havoc in terms of wave breaking etc. can be witnessed on the ocean surface, seismic disturbances traversing as body waves traverse reflecting and refracting in the interior of the Earth along with surface waves (propagating on the curved surface of the Earth) converge at the antipode to create destruction.  Our studies here show the importance of choosing closed targets which aid in the process of repeated energy transfer to particles and often their thermalization. The waves keep propagating in the closed confines rather than getting dissipated over an extended region as would happen for extended targets. 

A mechanism for enhanced laser energy absorption in plasma microglobules is demonstrated using 2D PIC simulations. The excitation and repeated collisions of surface and bulk waves within the confined target, lead to episodic energy surges and efficient thermalization. The microglobule size determines absorption efficiency and the timing of energy enhancements. Unlike extended targets, closed plasma targets confine waves, enabling sustained and maximized energy transfer

%It is shown that as the laser hits the overdense  micro-droplet target, electrons are extrated from the surface. These form  electron density waves and  propagate over the surface of the micro-droplet in opposite directions. However, ultimately they collide at the diametrically opposite point of the target resulting in the breaking of the waves and irreversible energy transfer to electrons. This obviously  can not happen for planar targets. In fact, thereafter, the remnant of the waves again propagate in opposite direction along the curved surface of the target and suffer repeated collisions when they collide.  Even a  bulk wave as it sloshes inside the target is observed to hit  these surface waves. During every collision the surface waves break and  a spurt of energetic electrons are produced.  The timing of  bursts in energy absorption has been shown to be correlated with these collisional events, thereby, providing evidence for the important role played by the colliding surface waves in the process of energy transfer. This work  touches upon the fundamental question of irreversible energy transfer from waves to particles where one often relies on wave breaking process.  The study here  demonstrates that even if the conditions for wave breaking are not not met, a viable alternative will be to devise schemes wherein the waves can collide and break. A possible experiment along the lines has also been proposed.

\end{abstract}

%\end{document}
                              %display desiredhttps://www.overleaf.com/project/64ac0527189a5be8c9cf30c2
\maketitle
%\tableofcontents
%\section{\label{sec:Introduction}Introduction}
%\textit{Introduction.-}
%\end{document}

\section{Introduction} 
\label{introduction}
The process of irreversible energy transfer from the laser Electromagnetic field (EM) to particles  in a plasma has intrigued researchers for a long time. Collisional processes, resonance mechanisms and conversion to electrostatic waves which subsequently transfer energy to particles via wave breaking and/or phase mixing, have often been invoked for this purpose \cite{PhysRevLett.74.1355,PhysRevE.47.3585,PhysRevLett.110.215002,PhysRevE.63.026406,PhysRevLett.61.90, katsouleas1988wave,sengupta2011phase,bag2024fluid}.  There have also been attempts to devise ways and means to enhance energy absorption, seek localized absorption, and/or transfer energy predominantly to specific plasma species, etc. \cite{maity2022mode,juneja2023ion,Vashistha_2020,vashistha2023localized,kaw2017nonlinear,das2020laser,nishida1987high,gibbon1996short,pukhov1999particle,brunel1987not,kruer2019physics,gershman1962propagation,kim1974development,arber2015contemporary,boyd1986trouble,ebrahim1980hot}.   Apart from deep fundamental interest,  this question is also very relevant for applications in electron/ion acceleration, designing high brightness high energy electromagnetic sources and laser fusion \cite{lindl1974,lindl1975two,lindl1974report,mikaelian1984density,christopherson2019thermonuclear}.  In this work, we illustrate a novel mechanism that is operative in the context of small finite-sized plasmas which we term here as plasma micro-globules. A 2-D particle-in-cell (PIC) simulation has been carried out for this purpose which shows that the colliding surface and bulk waves, triggered by laser, easily succumb to wave breaking and efficiently transfer energy to the particles. Notably such a scenario of colliding waves leading to a catastrophe is prevalent in a number of contexts cutting across domains.  For instance, ocean waves  crash on the shore and the reflected remnant (returning) wave  collides with the next  incoming wave. Whenever such a  collision happens at  an appropriate  phase, wave breaking occurs generating huge white water and  foam.  Another example is that of the bulk and surface waves triggered by earthquake which traverse the earth from different paths and converge at antipodes where they may wreak havoc.  In this study we show that similar processes  in micro-globular plasmas  facilitate efficient energy transfer in comparison to conventional planar targets.  Note that this study opens up a regime of laser absorption  very different from  the much  investigated and well understood mechanisms like  resonance absorption and vacuum heating \cite{brunel1987not}. Our study should have immediate and significant relevance for  plasma  acceleration and particle radiation  sources (used for imaging and semiconductor and  lithography) that are driven by kilohertz and multi-kilohertz  femtosecond lasers presently limited to sub relativistic intensities \cite{gamaly2011femtosecond,gamaly2022femtosecond}

This paper is organized as follows.  In Section II, we describe the simulation set-up. Section III contains diagnostics and analysis that illustrate the novel aspect in the globular plasma target that leds to enhanced laser absorption . Section IV summarizes  and concludes the study.

% We now describe the simulation set-up which is followed up with diagnostics and analysis that illustrate the novel aspect in the globular plasma target that leads to enhanced laser absorption. We finally summarize and conclude the study. 

%We provide the details about the simulation, diagnostics, and analysis below. 

\section{Simulation Details}
\begin{figure}
    \centering
    \includegraphics[width=1\columnwidth,keepaspectratio]{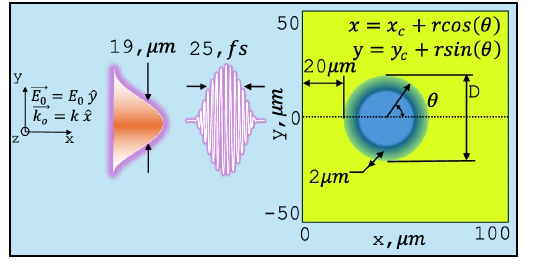}
    \caption{The figure illustrates a schematic drawing of the simulation box. The light green box inside the figure indicates the simulation domain. The laser pulse polarized along $\hat{y}$ is incident from the left side.}
    \label{Fig:schematic}
\end{figure}

%We provide the details of the simulation set up here. 
A 2D particle-in-cell (PIC)  simulations have been carried out to illustrate a novel mechanism that is operative in the interaction of laser with a finite-size plasma micro-globule.  The fully relativistic, massively-parallel PIC codes, EPOCH and OSIRIS  has been used for this purpose\cite{arber2015contemporary,fonseca2002osiris}. The 2D schematic of the 
 X-Y plane has been depicted in Fig.{\ref{Fig:schematic}}. The simulation box is $100 \mu m \times 100 \mu m$ with $12000 \times 12000$ cells and the grid size is $dx = dy = 0.00833 \mu m$. A fully ionized electron-ion plasma in the shape of a circular microglobule is placed inside the simulation box as shown in Fig.{\ref{Fig:schematic}}.  A laser pulse propagating along the $x$ axis is incident normal to this plasma globule. 

The ion-electron mass ratio is taken as  $1836$. Other details about the simulation parameters have been shown in Table \ref{table1}. We consider three distinct values of the diameter of the plasma globular target in our studies, namely  $15 \mu m$, $30 \mu m$, and $60 \mu m$. For comparison we consider the case of a  planar target.  The microglobular target  has a density gradient extending $2 \mu m $ from its boundary towards the interior. The case of a sharp density profile is also considered. The density in the interior is constant and is about $10$ times the critical density of the laser.  The laser pulse has a wavelength of $800 nm$ and a spot size with full-width at a half-maximum (FWHM) is $19 \mu m$.  The pulse duration is about  $25 fs$. The laser intensity for these studies has been restricted to non-relativistic regime where the normalized vector potential has a value of $a_{0}=0.5$. As mentioned earlier, the laser propagates along $x$ direction and hits the vacuum-plasma interface from the left side of the simulation box. The laser is chosen to be plane polarized with its electric field ($E_{l}$)  along the $y$ axis, as shown in Fig. \ref{Fig:schematic}. The boundary conditions are taken as absorbing for fields as well as for particles.
%In this work 2D PIC simulation have been carried out to study ultrafast nonlrelativisitic laser interaction with overdense plasma. Fully relativistic massively parallel PIC code EPOCH [ref] has been used. The 2D Simulation domain is in X-Y plane. The overdense circular plasma is centered on y=0 and the front point of the circle is 20 microns from the left boundary from which the laser is launched. The number of cells as 12000 \times 12000. The grid size $dy=dx=1/60$ \mu $=m$.
The number of particles in each cell is chosen as  $8 \times 8$. The dynamics of both electrons and ions have been considered. 
%The mass of the ions is taken to be approximately the same as that of the proton, i.e.  $1836$ times the mass of electrons. 
%The initial background plasma temperature is chosen to be $0.025 eV$. The laser is launched from the left boundary with propagation vector $\vec{k}$ along $\hat{x}$, incident on the vacuum plasma interface. The laser is p-polarised along $\hat{y}$. The temporal width is considered to be Gaussian with a full-width half maximum of 25 fs. The intensity of the laser pulse is  $$.... W/cm^2$ corresponding $a_0 = 5$. The simulation parameters are presented in Table 1 [ref]
\normalsize
    \begin{table}
    \caption{\label{table1} Simulation Parameters}
    \begin{center}
    \begin{tabular}{l l l }
    \hline
    \hline
    Parameters                 &                   &Standard Units  \\
    \hline
    \hline
    \textit{Laser}  \\
    Wavelength                & $\lambda_{L}$     & 800 nm \\
    Frequency                 & $\omega_{L}$      & 3 \\
    Intensity                 & $a_{o} $     & $0.5$, $5.37\times10^{17} W/cm^{2}$ \\
    \hline
    \textit{Plasma} \\
    Electron Plasma frequency & $\omega_{pe}$     & $7.443 \times 10^{15} rad/s$\\
    Critical density          & $n_{c}$& $1.74 \times 10^{27} 1/m^{3}$\\
    Electron number density   & $n_{e}$           &  $10 n_{c}, 1.74 \times 10^{28} 1/m^{3}$\\
    Ion number density        & $n_{i}$           & $1.74 \times 10^{28} 1/m^{3}$\\
    Mass of electrons         & $m_e$             & $9.109 \times 10^{-31} kg $ \\
    Mass of ions              & $m_i$             & $1836  m_{e}$ ,$ 1.677 \times 10^{-27}kg$ \\
    \hline
    \hline
    \end{tabular}
    \end{center}
     \end{table}
A preliminary  3-D simulation study carried out  confirms the salient observations and inferences drawn by 2-D studies. 

\section{Observations}
\begin{figure}
    \centering
    \includegraphics[width=1\columnwidth,keepaspectratio]{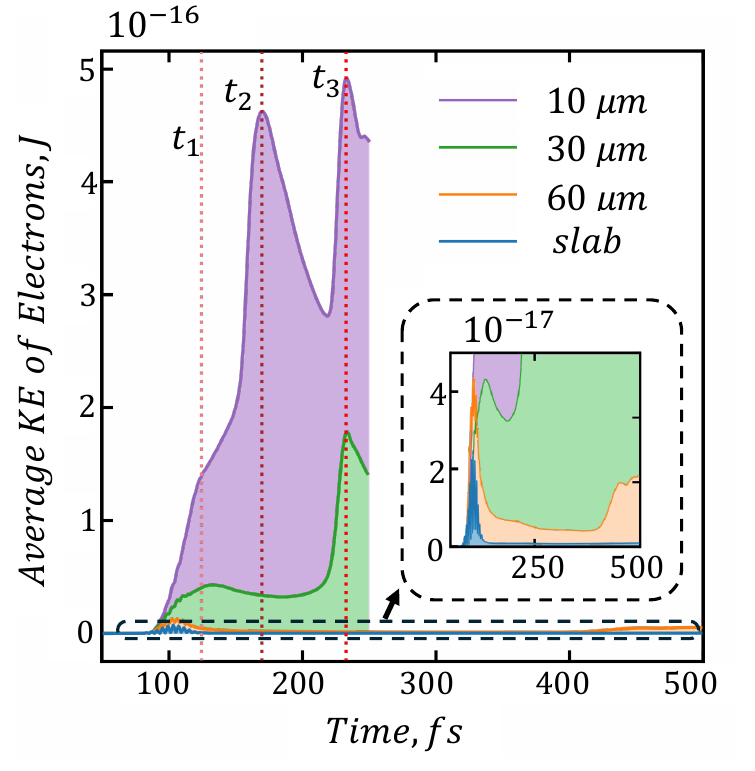}
    \caption{ Temporal Evolution of Average Kinetic Energy of Electrons for target diameters of $15 \mu m $, $30 \mu m$,$60 \mu m $ and $slab$. The peaks at $t_1 , t_2 $ and $ t_3$ in $15 \mu m $ target correspond to the time of the hump (for ease we call it a first peak) and the appearance of the two peaks respectively. Clearly, at these times some episodic phenomena occur which facilitates a large transfer of energy.  }
    \label{Fig:KE-ele-Time}
\end{figure}

\begin{figure*}
    \centering
    \includegraphics[width=2.05\columnwidth,keepaspectratio]{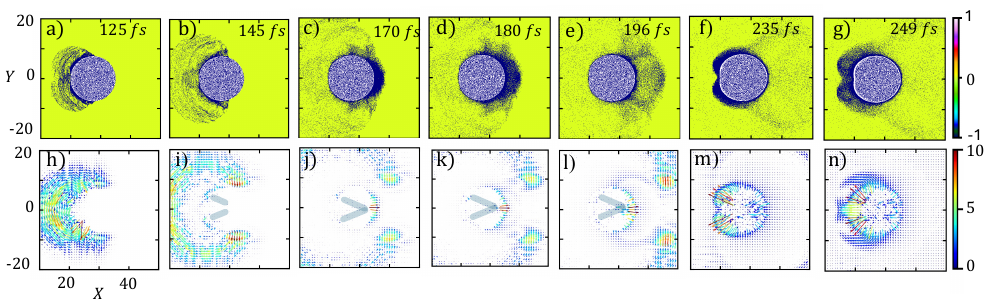}
    \caption{Snapshots of Charge density (a-g) and Electric Field vector (h-i) for $15 \mu m $  target. The electrons are blue and ions are white. The electrons are pulled out of the surface in the region of laser incidence. Disturbances in electron density are seen propagating forward and eventually colliding at subplot(d). The stochastic bulk waves are seen propagating through the bulk in the quiver plot of the electric field. Enhancements in the field are seen where the surface and bulk disturbances collide}
    \label{Fig:dens-field-time}
\end{figure*}

%As shown in the schematic plot of Fig.1,  a laser with non-relativistic intensity is incident normal to the circular overdense target from the left side. 
 The evolution of electron kinetic energy (KE) in the target is shown in Fig.\ref{Fig:KE-ele-Time}. The laser hits the plasma target at $t = 66 fs$, following which  the KE increases.  This figure shows  KE evolution for  three different  values of the target diameter. The acquired KE is highest for the smallest sized $15 \mu m$ target. For the planar target there is negligible energy gain even when the simulation is run for a much longer duration. The other characteristic feature of these energy gain plots is the appearance of distinct peaks. For the $15 \mu m$ target the timings of the three peaks have been depicted by $t_1, t_2$ and $t_3$ respectively. Increasing the diameter of the target  increases the interval between the peaks (see the inset).   These peaks indicate that a specific mechanism which enhances the energy transfer process from fields to particles, is operative in the interaction.  This work aims to unravel this novel mechanism of enhanced laser energy absorption process.  We emphasize that the  mechanism is operative only in the context of small finite-sized globular plasma targets which offer a confined region. 
We depict the spatial distribution of the charge density and the electric field quiver plots for the smallest $15 \mu m $ target in Fig.\ref{Fig:dens-field-time}. Some of these snapshots have been taken at the intervening times $t_1, t_2 $ and $t_3$  demarcated by three vertical lines in Fig.\ref{Fig:KE-ele-Time} about which we discussed in the previous paragraph. Additional snapshots taken  at intervening times. The complete evolution as a movie is placed as supplementary material [] 
%The complete evolution of the same has been provided as a movie in the attached supplementary material to this article. 

We now  focus on the plots of Fig.\ref{Fig:dens-field-time} corresponding to $t = t_1$.  The kinetic energy of electrons increases and forms the first peak of Fig.\ref{Fig:KE-ele-Time}. The spatial distribution of electrons shown in Fig. \ref{Fig:dens-field-time} at $t=124 fs$, clearly shows that the electrons are extracted by the laser into the vacuum region on the left side of the target. The extraction is maximum at a certain oblique angle $\theta=113 ^ {o}$ (depicted in Fig.\ref{Fig:schematic}) forming two bulges in the left surface of the target for    $ 90^{o} < \theta <  270^{o}$. At $\theta = 180^{o}$ there is a dimple in the spatial electron distribution. The electrons cannot be extracted from here as the electric field is tangential to the target surface.  This phase, where two lobes of electrons are extracted, depicted in Fig. \ref{Fig:dens-field-time}a,  occurs while the laser radiation is present and interacting with the left side of the target surface ($ 90^{o} < \theta <  270^{o}$).  The energy absorption in this particular phase is essentially due to  Brunel heating  \cite{brunel1987not}. The laser electric field which is normal at $\theta = 180 ^{o}$ becomes oblique at other angles due to the curvature of the target. It is for this reason that no such electron extraction occurs for planar targets.  It is also evident from Fig.\ref{Fig:KE-ele-Time} which shows that the kinetic energy acquired by the electrons for the planar target is negligible.  
%It should be noted that the curvature of the plasma microglobular target is comparable to the laser waist. 
%Hence, the angle at which the laser hits the target is oblique and varies with the surface locations.  The vacuum heating criteria of having a component of the laser electric field normal to the target surface is thus satisfied as the various locations of the target surface. The dimple in spatial electron distribution in a vacuum at $y =0$ (the target center)  and the up-down symmetry of the emitted electrons from the surface can thus easily be understood.  Thus, the first peak essentially in the kinetic energy evolution occurs due to the vacuum heating process enabled by the curvature of the target. 
The laser impinges upon the target surface with increasing values of $\alpha = \mid \theta - 180^{o}\mid$ (from zero to $90^{o}$).   A surface disturbance  of electron density adhering to  and propagating along  the target surface is  observed during this phase.  After  hitting the target at the top ($T$)  and bottom ($B$) surfaces at $\alpha = 90^{o}$, the laser simply grazes past the target. The right side of the target is thus not illuminated by the laser. The surface wave of the electron density, on the other hand,   propagates along the curvature of the target surface from both top and bottom sides and collides at $\theta = 0^{o}$ when $t = t_2$ corresponding to the appearance of the peak shown in  Fig.\ref{Fig:KE-ele-Time}. For a target with a bigger diameter this surface wave takes a longer time to arrive and collide at $\theta = 0^{o}$. This explains the increased duration between peaks with increasing diameter of the target.  This  matches with simple estimates as we now demonstrate.  
For the 15 $\mu m$ target the interval between collisions $(t_2-t_1) = \Delta t_{15} \sim 48 fs$. For the target with double the diameter  ($30 \mu m$) this interval  is $ \Delta t_{30} \sim 101 fs$. We note that the ratio of the two intervals $\Delta t_{15}/ \Delta t_{30 } \sim 48/101 \approx 2 = 15/30 $. For bigger targets, as expected the disturbances take proportionately longer time to travel from opposite sides and collide .  % reach from the front region where the electron is ejected to propagate forward and collide at $R$ and an additional $62.02 fs$ for the waves to propagate along $-X$ direction and collide in the front $F$ region. In the case of the 30-micron target, the collision occurs after a time interval of $101 fs$ whereas the maximum time taken for the waves to collide is $241 fs$ for the case of the $60 \mu m$ target}. 

%{\textcolor{teal}{The surface wave propogates $\sim c$. For the 15 \micron target it takes $\sim 78 fs $  to for the collision to take place at $\theta=0^{0}$. when the size of the target is doubled, the collisions take place after $157.2 fs$. and for 60 micron target it rakes $\sim 313 fs$ for the waves to reach $\theta = 0^{0}$}

For the $15 \mu m$ target the time $t_3$ refers to the time  when the disturbance sloshes back and hits the front surface of the target as can be observed from the subplot (f) of Fig.\ref{Fig:dens-field-time}.  The  electric field quiver subplots in this figure  also show a surface wave propagating back and forth. In addition, there is clear evidence  of a bulk wave propagating inside the target. This can be observed from the highlighted region in the electric field quiver plots. It is essentially a stochastic electric field that sloshes back and forth \cite{puri1966plasma} in the bulk region and carries energy with itself. The stochastic nature is evident from the random direction of the electric field in the highlighted region.
\begin{figure*}
    \centering
    \includegraphics[width=2\columnwidth]{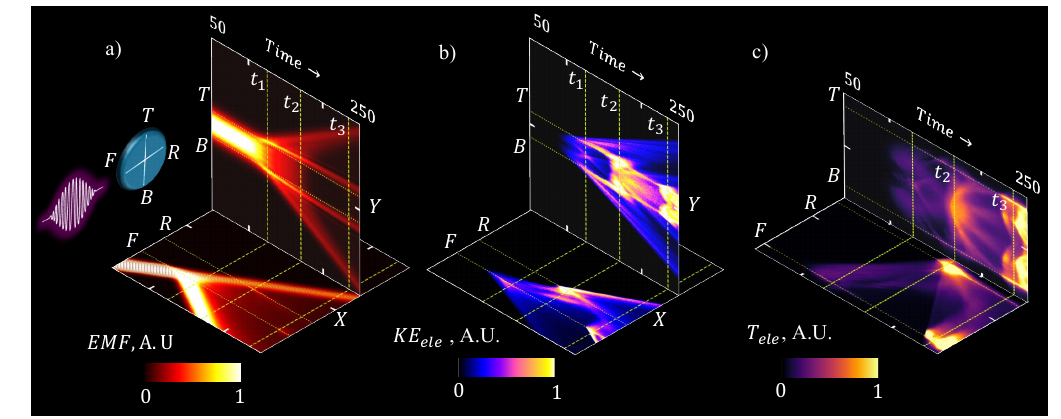}
    \caption{The electromagnetic field energy, kinetic energy of electrons and their thermal energy has been shown  in the space time plane  in subplot (a), (b) and (c) respectively. The bright colored regions depict higher value of these fields.  The letters F,R, T, B denote  front, rear, top and bottom position of the target.  The reflection of Electromagnetic field energy from  F  and propagation forward along can be clearly seen in the plot in $X $ vs. $t$ plane in subplot (a). The times stamps $t_1, t_2$ and $t_3$ are depicted by straight lines in the time axis.}
    \label{Fig:space-time}
\end{figure*}
The signatures of the surface wave and the bulk wave propagation has been further demonstrated clearly in  Fig.\ref{Fig:space-time}. This figure shows the space-time plot for the electromagnetic field energy (subplot(a)), and the kinetic (subplot(b))  and thermal energy (subplot(c)) acquired by the electrons. The evolution of these has been shown in the  $X vs. t  $ and $Y  vs.  t $ planes. Here $t$ represents the time. In the plane $X  vs.  t$ and $Y  vs.  t$ plots these quantities have been integrated along $Y$ and $X$ respectively. 
As the thermal energy shows up primarily only near  the target region, the space (both the $X$ and $Y$ axis)  has been zoomed  for a better perspective in subplot(c). The plots represents the data of the target size  $15 \mu m$. 
%Simulation data for both $15 \mu m$ and $30 \mu m$ diameter plasma microglobules have been presented. In the X-time plane, the quantity of interest is averaged along the Y direction, and in the Y-time plane, they are averaged along the X-axis. 
The  F(front), R(rear)  Top(T), and Bottom(B) of the target location have been demarcated by straight lines in the space-time plots. The time axis has also been demarcated at three locations corresponding to the appearance of the peak in the kinetic energy plot of Fig.\ref{Fig:KE-ele-Time}.  As the laser falls on the front surface of the target it  primarily gets reflected from the  overdense plasma. The bright broad horizontal strip in the $Y vs. t$ plot on subplot (a) denotes the incoming and reflected laser. In the $X  vs. t$ plot, the forward propagation in $X$ and the reversal in the $X$ axis clearly shows the  reflection of the  EM fields along $X$.  A  part of the EM field, however,  continues to propagate forward in $X$ circumventing the target surface (see subplot(i,j,k,l) of Fig.\ref{Fig:dens-field-time}. In the $Y  vs. t$ plot, this can be seen as two rays propagating along positive and negative $Y$ with time. Furthermore, two more rays of  EM wave energy can be observed to graze past the top and bottom surfaces of the target. %This, however, alternates in time in the $X-t$ plane.
%In fact, this illustrates that there is an electromagnetic field  that clings to the target surface. This in fact is associated with the surface wave motion illustrated in Fig. 3. 
The plots of kinetic energy and thermal energy in  Fig.\ref{Fig:space-time}(b) and Fig.\ref{Fig:space-time}(c) clearly provide evidence for kinetic and thermal energy enhancements. In fact these enhancements occur predominantly when the laser has already  left the simulation box.  The timings of the enhancements of these energies are same as those of the collisions of disturbances  at the rear and again at the front surface. In addition in both these energy plots the enhanced brightness,  within the lines denoting the front-rear and top-bottom surface of the target,  is an indicator of the bulk disturbance. This bulk wave  evidently emanates from the two bulging regions of the front surface of the target and converge at the other side of the target.  A careful inspection  of the plots of Fig.\ref{Fig:space-time} in fact confirms our earlier assertions on episodic increase of particle energy when the surface and bulk waves collide.

It should thus be noted that the irreversible energy gain (both kinetic and thermal)  by electrons occurs repeatedly as the disturbances continue to collide alternatively at the front and rear surfaces. This keeps occurring even after the laser has left. The  planar slab  target does not provide a confined domain for the waves to collide repeatedly. The finite region of the micro-globular targets constrain these waves and offers the opportunity for repeated collisions between them facilitating an irreversible and efficient process of energy transfer.   
% \begin{figure}
%     \centering
%     \includegraphics[]{Figures/ekSpec.png}
%     \caption{Evolution of Electron energy spectrum (t1, t2, t3)}
%     \label{fig:ene-ele-spec}
% \end{figure}

% \begin{figure}
%     \centering
%     \includegraphics[]{Figures/Fig-angdis-dens-theta.pdf}
%     \caption{Anglular distirubuiton (a,b) and charge density at $7.5 \mu m$ c,d at times $t_a and t_b$}
%     \label{fig:enter-label}
% \end{figure}
\begin{figure}
    \centering
    \includegraphics[width=0.86\linewidth]{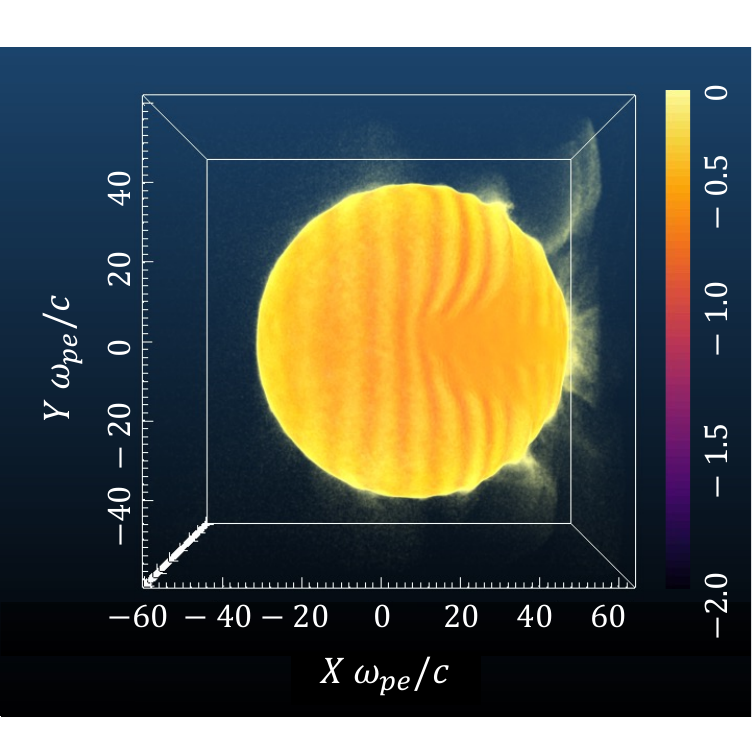}
        \caption{Snapshot of Normalized Electron Density of $15 \mu m $ target showing the surface waves and ejection of electrons at rear side of the the target.}
    \label{fig:run3d-elel}
\end{figure}

% We also show the  energy spectrum of  the electrons as a function time Fig. \ref{fig:ene-ele-spec} . This clearly shows that the even at these non relativistic intensity of the laser  the electrons acquire energy of the  order of  MeV. The angular distribution of this energetic electrons have been shown in Fig. \ref{fig:angdis-dens-theta}
%  As expected from the front side the electrons get emitted around  two oblique angles $\theta = \pm .. $ while at the back side the emission is predominantly around $\theta = 0$. 

% \textcolor{teal}{the the percentage of laser energy absorbed by $60 \mu m$} is $3.9\% $, for $\% micron target is $ 

We have carried out a wide variety of  simulations with varying conditions to confirm our conclusions.  For instance, we have studied oblique incidence of the  laser on the micro-globules.   In this case the disturbances that get generated are asymmetrical. When the two disturbances collide, the one with  higher amplitude is seen to  drag the smaller one around the surface of the target. The energy transfer process in asymmetric cases gets  reduced. Furthermore 3-D simulations with spherical targets have also been performed. A typical snapshot has been shown in  Fig.\ref{fig:run3d-elel}.  The surface disturbances and their propagation on the spherical surface is  clearly visible in this figure.  These  waves are observed to converge and lead to an irreversible energy transfer to electrons.

Let us also reflect on the suitable micro-globule target  size for which this mechanism will be most efficient. We have already demonstrated through the simulation of the various  target sizes  that  the  larger the size 
 the smaller  the energy gain. The lower limit of the target size gets decided by the laser wavelength. If the target size is chosen to be smaller than the laser wavelength then the electron cloud in the entire globule   merely oscillates around the ion background as a single entity. Thus the wave like disturbances , are not generated. 

\section{Conclusions}
We show that the the efficiency of laser energy conversion into  the kinetic and thermal energy of electrons can be dramatically increased by choosing a micro-globular target. This happens as the finite closed region of the micro-globular target constrains the disturbances and permits their repeated collisions releasing field energy to irreversible particle kinetic energy. Similar  occurrence of energy release through colliding disturbances in closed regions has been viewed in the context of earthquake. The surface and body disturbances in this case propagate on the surface of the earth or in the interior and converge at antipodes to wreck havoc there. The message that our work here conveys is that  using similar construction one can  extract and convert laser energy efficiently into particle kinetic energy.

\section{Acknowledgements}
The authors thank IIT Delhi HPC facility for providing
computational resources. AD acknowledges support from the  Anusandhan National Research Foundation (ANRF) of the Government of India through core grant CRG/2022/002782 as well as a  J C Bose Fellowship grant JCB/2017/000055. GRK thanks the Department of Atomic Energy for long term support of this research  and ANRF for a J C Bose Fellowship grant JBR/2020/000039. AS Acknowledges support from Anusandhan National Research Foundation (ANRF) of the Government of India through core grant CRG/2022/002782.

\bibliographystyle{apsrev4-1}
\bibliography{main}% Produces the bibliography via BibTeX.

\end{document}